\documentclass[useAMS,usenatbib]{mn2e}

\usepackage{graphicx}
\usepackage{url}
\usepackage{float}
\usepackage{color}

\usepackage{txfonts}

\usepackage{booktabs}

\usepackage{enumerate}

\title[GTC observations of RWT\,152]{Observations of the planetary nebula RWT\,152 with 
OSIRIS/GTC\thanks{Based on observations made with the Gran Telescopio Canarias (GTC), installed
at the Spanish Observatorio de El Roque de los Muchachos of the Instituto de Astrof\'{\i}sica
de Canarias, in the island of La Palma.}}
\author[A. Aller et al.]{A. Aller$^{1,2,3}$\thanks{E-mail: alba.aller@ifa.uv.cl (AA)}, L. F. Miranda$^{4}$,
L. Olgu\'{i}n$^{5}$, E. Solano$^{2,6}$ and A. Ulla$^{3,7}$\\
$^{1}$Instituto de F\'isica y Astronom\'ia, Facultad de Ciencias, Universidad de Valpara\'iso, Gran Breta\~na 1111, Playa Ancha, Valpara\'iso, 2360102, Chile\\
$^{2}$Departamento de Astrof\'{\i}sica, Centro de Astrobiolog\'{\i}a
(INTA-CSIC), PO Box 78, Villanueva de la Ca\~nada (Madrid) E-28691, Spain\\
$^{3}$Departamento de F\'isica Aplicada, Universidade de Vigo, Campus Lagoas-Marcosende s/n, Vigo E-36310, Spain\\
$^{4}$Instituto de Astrof\'{\i}sica de Andaluc\'{\i}a - CSIC, C/ Glorieta de
la Astronom\'{i}a s/n, E-18008 Granada, Spain\\
$^{5}$Departamento de Investigaci\'on en F\'{\i}sica, Universidad de Sonora,
Blvd. Rosales Esq. L.D. Colosio, Edif. 3H, 83190 Hermosillo, Son. Mexico\\     
$^{6}$Spanish virtual observatory, PO Box 78, Villanueva de la Ca\~nada (Madrid) E-28691, Spain\\ 
$^{7}$Astronomy Unit, Queen Mary University of London, Mile End Road, London E1 4NS, UK\\   
}
\begin{document}

\date{Accepted 1988 December 15. Received 1988 December 14; in original form 1988 October 11}

\pagerange{\pageref{firstpage}--\pageref{lastpage}} \pubyear{2002}

\maketitle

\label{firstpage}

\begin{abstract}
 RWT\,152 is one of the few known planetary nebulae with an sdO central star. We
present subarcsecond red tunable filter H$\alpha$ imaging and
intermediate-resolution, long-slit spectroscopy of RWT\,152 obtained 
with OSIRIS/GTC with the goal of analyzing its properties. The
H$\alpha$ image reveals a bipolar nebula with a bright equatorial region and
multiple bubbles in the main lobes. A faint circular halo surrounds the main
nebula. The nebular spectra reveal a very low-excitation nebula with weak
emission lines from H$^+$, He$^{+}$, and double-ionized metals, and 
absence of emission lines from neutral and single-ionized metals, except for an extremely 
faint [N\,{\sc ii}] $\lambda$6584 emission line. These spectra may be
explained if RWT\,152 is a density-bounded planetary nebula. Low nebular chemical abundances of
S, O, Ar, N, and Ne are obtained in RWT\,152, which, together with the derived high peculiar
velocity ($\sim$ 92-131 km\,s$^{-1}$), indicate that this object is a halo planetary nebula. The 
available data are consistent with RWT\,152 evolving from a low-mass progenitor ($\sim$
1\,M$_{\odot}$) formed in a metal-poor environment. 
\end{abstract}


\section{Introduction}

The formation of a planetary nebula (PN) is the last phase in the evolution
of low- and intermediate-mass stars ( initially 0.8-8 M$_{\odot}$) before
they become white dwarfs. PNe exhibit a wide variety of morphologies and
properties that should be closely related to the evolution of their central
stars (CSs). PNe host a diversity of CSs. \cite{Weidmann-Gamen2011} compiled 26
spectroscopically well-distinguished types of CSs, that include, for example,
hybrid stars, O-type, and hot subdwarf stars (sdOs and sdBs). However, there
are only a few PNe known that host an sdO CS. \cite{Aller2015a} (hereafter AM15),
list 18 PNe around sdOs CSs. Although this number is probably
a lower limit because several CSs still lack a firm classification as, e.g.,
NGC\,1514 \citep{Aller2015b} and NGC\,6026 \citep{DeMarco2009, Hillwig2010}, 
the number of PN+sdO systems is very small as compared with the
large number of known sdOs without associated PNe (see, e.g., \citealt{Ostensen2006})
and with the number of known Galactic PNe \citep[$\sim$ 3000,][]{Frew-Parker2010}. 
 Intriguingly, PNe around sdO CSs share some characteristics: they are
generally evolved or relatively evolved PNe ($\sim$ 10$^{4}$ yr) with a very low surface brightness; most of them
present elliptical and bipolar shapes, often with multiple structures,
including signs of collimated outflows; and a high fraction of confirmed (or suspected)
binary CSs is also observed in these PN+sdO systems (for more details see AM15 and references therein).

The existence of these common characteristics strongly suggests a common formation process for
these PNe as well. On the other hand, the small number of these systems points out that 
some peculiarities may be present in their formation. Among the
various evolutionary paths to explain the formation of an sdO star \citep[see][]{Heber2009}, 
 post-asymptotic giant branch (post-AGB) evolution appears as the most suitable one for those 
sdOs with PNe around them. If so, the small number of these systems implies that PN ejection is a rare evolutionary path for sdOs. 
However, as sdOs are thought to represent the late stage of low-mass stars \citep[see, e.g.,][]{Heber2009}, 
whose evolution proceeds very slow, the number of PN+sdO systems could be biased due, at least in part, to the fact that the shell ejected during
the post-AGB phase could have dispersed before being photoionized. In any case, PN+sdO
systems are peculiar and their origin should be investigated. An interesting
approximation to study these systems is analyzing the properties of the PNe themselves 
(e.g., morphology, chemical abundances), which may provide clues about their formation and evolution.

RWT\,152 ($\alpha_{(2000.0)}$ = 07$^h$\,29$^m$\,58$\fs$5, $\delta_{(2000.0)}$
= $-$02$^{\circ}$\,06$'$\,37$''$; {\it l} = 219$\fdg$2, {\it b} = 7$\fdg$5) is one of the few PNe hosting a CS
classified as sdO \citep{Ebbets-Savage1982}. This compact
and faint PN was originally discovered by \cite{Pritchet1984} and has recently
been analyzed by AM15 by means of narrow-band imaging, and low- and high-resolution long-slit
optical spectra. The images by AM15 show a faint bipolar PN in the light of [O\,{\sc
  iii}] while a diffuse, non-spherical shape can be distinguished in the light
of H$\alpha$. The analysis of the internal kinematics by AM15 showed a basic pattern of bipolar motions, 
although these authors noticed some deviations
from a pure bipolar (hour-glass) shell, suggesting a more complex
morpho-kinematic structure. In addition, the polar velocity of 
19\,km\,s$^{-1}$ derived by AM15 and the published distances of 1.4 and 6.5\,kpc 
\citep[respectively]{Ebbets-Savage1982, Pritchet1984} to RWT\,152,
yield a kinematical age of $\simeq$ 6.6 and 17.8 $\times$10$^3$\,yr, respectively, which
suggests an already evolved PN. Finally, only H$\alpha$, H$\beta$, and [O\,{\sc
  iii}]$\lambda$$\lambda$4959,5007 emission lines were detected in the nebular
spectra, which leaded AM15 to suggest a possible deficiency of 
heavy elements in the nebula.
Confirmation of this possible deficiency requires much deeper spectra given
the faintness of RWT\,152. In addition, images at higher spatial resolution
are necessary to analyze its morphology in detail. In this framework, large
telescopes represent a very useful tool to carry out detailed analyses of such
faint PNe, since they provide very high-quality data with relatively
short-exposure times. 

In this work we present subarcsecond red tunable filter imaging and deep
intermediate-resolution, long-slit spectra of RWT\,152 
obtained with OSIRIS/GTC, with the aim of studying the morphology,
physical conditions and chemical abundances of this PN. The images reveal new
structures not observed so far while the spectra allow us to detect very faint nebular 
emission lines not previously detected, allowing us estimates of the chemical
abundances in the nebula. The layout of the paper is
as follows: in Sect. \ref{Section:observations} we summarize the
observations and data reduction. Sect. \ref{Section:results} describes the
main results derived from the imaging and intermediate-resolution 
spectra. A discussion of the results is done in Sect. \ref{Section:discussion}
and the main conclusions are summarized in Sect. \ref{Section:conclusions}.

\section[]{Observations and data reduction}
\label{Section:observations}

\subsection{Optical imaging}

Images of RWT\,152 were obtained on 2013 November 6 in service mode (proposal ID GTC4-13B)
with OSIRIS (Optical System for Imaging and low-Intermediate-Resolution Integrated Spectroscopy), 
 mounted on the Nasmyth-B focus of the 10.4\,m Gran Telescopio Canarias (GTC), at the Observatorio Roque de los Muchachos 
(La Palma, Canary Islands, Spain).
The detector of OSIRIS consists of two Marconi 
CCD42-82 (2048$\times$4096 pixels) with a 37 pix (binned) gap between them. The maximum 
unvignetted field of view (FOV) is 7.8$\times$7.8 arcmin. In order to increase the signal-to-noise 
we chose the standard 2$\times$2 binning mode which provides a 
plate scale of 0.254 \,arcsec pixel$^{-1}$.

  \begin{figure}
\includegraphics[width=0.48\textwidth]{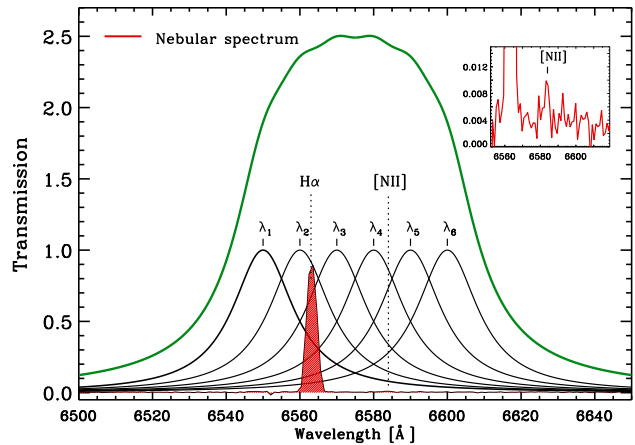}
  \vspace*{1pt}
  \caption{ Transmission curves of the OS filter normalized at the peak transmission for the six selected bands 
(labelled as $\lambda$$_{1}$, $\lambda$$_{2}$, $\lambda$$_{3}$, etc.) are plotted in black. 
  The scaled nebular spectrum of RWT\,152 (see Sect. 3.2) is overimposed in red and the position of the H$\alpha$ and 
  [N\,{\sc ii}] emission lines is indicated to highlight 
  the contribution of these lines to the different bands. The resulting band of adding each individual band is plotted 
  in green. The inset shows the spectrum in the range 6550--6620 \AA, in order to highlighted the weak 
  [N\,{\sc ii}]\,$\lambda$\,6584\,{\AA} emission line detected.}
\end{figure}

The red tunable filter \citep[RTF,][]{Cepa2003,Cepa2005} was used, which covers the
6510--9345\AA{} wavelength range. When using the RTF, the user must select one of the available 
Order Sorter (OS) filters in order to isolate the desired order. Thus, we selected the OS filter f657/35 
(i.e., that at central wavelength 6572\,$\AA$ and FWHM of 350\,$\AA$) that provides a wavelength 
range of 6490--6600\AA{}, with the aim of covering the H$\alpha$ and [N\,{\sc ii}]\,$\lambda\lambda$\,6548,6584\,{\AA}
emission lines from the nebula. The RTF was sintonized with a full width at half-maximum (FWHM) of 20 \AA{}.

It should be noted that with the RTF the wavelength along the FOV is not
uniform, decreasing radially outwards from the optical centre following the
law  
 
 \begin{equation}
 \lambda = \lambda_{0} - 5.04 \times r (arcmin)^{2},
\label{Eq:Flambda}
\end{equation}
where $\lambda_{0}$ is the central wavelength (in angstroms) and r the distance to the
optical centre \citep[see][]{Gonzalez2014}.

\begin{figure*}
\includegraphics[width=1.0\textwidth]{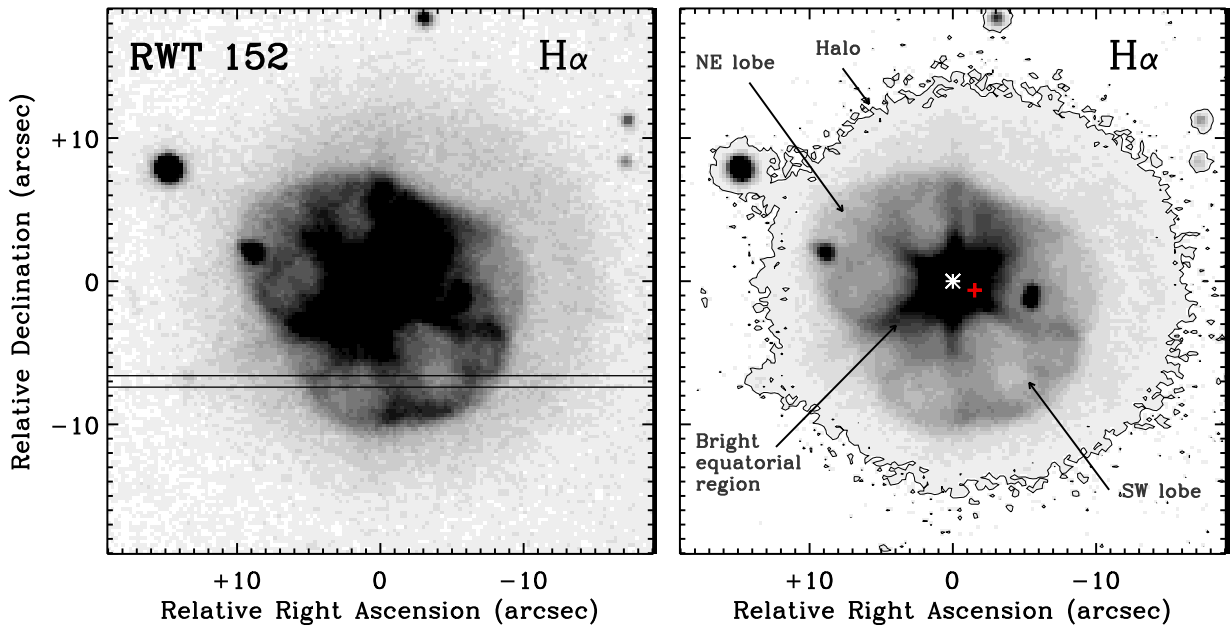}
  \vspace*{1pt}
  \caption{Grey-scale reproductions of the H$\alpha$ image of RWT\,152 (north is up; east to the left). Grey levels
are linear on the left-hand panel and logarithmic on the right-hand one.
The origin (0,0) is the position of the central star marked with a white
asterisk. The long slit used for spectroscopy is indicated by the horizontal lines
in the left-hand panel and the different features of the nebula are indicated in the right-hand panel (see text for more details). 
The contour in the right-hand panel traces
the size of the halo at the 3$\sigma$ level above the background; the
geometrical center of the halo is indicated by a red plus symbol.}
\end{figure*}

This quadratic decrease of the wavelength is particularly critical in extended objects such as PNe, 
since the light collected from each point of the nebula comes from different wavelengths
depending on the distance to the optical centre. Taking this into account, and as mentioned above, 
the observations were designed to 
cover the H$\alpha$ and [N\,{\sc ii}]$\lambda$6548,6584 emissions
from all the points of RWT\,152. The observing strategy was to obtain images with the
optical centre at six different central wavelengths ($\lambda_{0}$ = 6550,
 6560, 6570, 6580, 6590, and 6600 \AA{}), hereafter referred to as individual
 bands. Figure\,1 shows the transmission curves of the selected OS filter normalized at the peak transmission
 centred at these individual bands (black lines) and the positions of the H$\alpha$ and [N\,{\sc ii}]$\lambda$6584 
 emission lines are indicated. The intermediate-resolution,
 long-slit spectrum of RWT\,152 (see Sect.\,3.2) has also been plotted in Fig.\,1 to
 show the contribution of the H$\alpha$ and the possible (very faint) [N\,{\sc ii}]$\lambda$6584 emission line detected.
  We note that the observed flux of the
 [N\,{\sc ii}]$\lambda$6584 emission line is practically negligible, only
 $\sim$0.5\% of H$\alpha$ (see Sect.\,3.2), thus making the H$\alpha$ emission line the
 main contributor in this OS filter. The CS of RWT\,152 was placed at $\sim$
 1\,arcmin from the optical centre, corresponding to $\sim$ 5 \AA{} bluer than
 the central wavelength of each band (see equation 1). However, the small size
 of RWT\,152 (see Sect. 3.1) implies that the wavelength along the nebula
 hardly varies, so we can assume that all points of the nebula are covered by
 approximately the same wavelength. Therefore, for this particular case, by adding the images 
  of all individual bands we obtain an H$\alpha$ image of RWT\,152. Three
  images were taken in each individual band, each with an exposure time of
  80\,s, allowing a dithering of $\sim$5 arcsec between images to properly
  remove the diametric ghosts produced in the images.

The images were reduced using the standard procedures within {\sc iraf}\footnote{IRAF is distributed 
by the National Optical Astronomy Observatory, 
which is operated by the Association of Universities for Research in Astronomy (AURA) under a 
cooperative agreement with the National Science Foundation.}.
After correcting each individual frame from bias and flat-field, the three images of each 
individual band were aligned and median combined. Then, we added all images of
all individual bands to obtain a deep H$\alpha$ image within the total band
plotted as a green line in Fig\,1. The total exposure time for the final
H$\alpha$ image is 1440\,s (i.e., 6(bands)$\times$3(images)$\times$80\,s) and
the spatial resolution is 0.7\,arcsec, as indicated by the FWHM of field stars
in the image. Figure\,2 shows the H$\alpha$ image at two different grey levels. We note that 
the H$\alpha$ image contains some contribution of the nebular continuum (see Fig.\,1). Nevertheless, as we 
will see below, this contribution is very small and the final H$\alpha$ image in Fig\,1 can be 
considered as representative of the ``pure'' H$\alpha$ emission.

\subsection{Intermediate-resolution, long-slit optical spectra}
Intermediate-resolution, long-slit spectra of RWT\,152 were obtained with OSIRIS. The 
volume-phased holographic gratings (VPHs) R2500U, R2500V, R2500R, R2500I were used. 
They cover the spectral ranges 
3440--4610, 4500--6000, 5575--7685, and 7330-10000 \AA{}, respectively, at dispersions
 of 0.62, 0.80, 1.04, and 1.36\,$\AA$\,pixel$^{-1}$ (these dispersions are measured at central 
 wavelengths for a slit width of 0.6\,arcsec). The standard 2$\times$2 binning mode was used, 
  which provides a plate scale of 0.254 \,arcsec pixel$^{-1}$.

 Spectra were obtained on 2013 November 6 (R2500R, R2500I) and 7 (R2500U, R2500V). 
 The slit width was 0.8\,arcsec and the spectra were obtained with the slit oriented at PA 90$^{\circ}$ 
 and centred 7 arcsec south of the CS (see Fig.\,2, left panel), covering the southwestern lobe of the nebula.
The exposure time was 1200\,s for each VPH and the seeing was $\simeq$ 0.9\,arcsec.

The spectra were reduced following standard procedures for long-slit
spectroscopy within the {\sc iraf} and {\sc midas}\footnote{MIDAS is developed
  and maintained by the European Southern Observatory.} packages. The
reduction included cosmic rays removal, bias subtraction and flat-field
correction. Then, the spectra were wavelength calibrated, sky subtracted and,
finally, flux calibrated using the spectrophotometric standard Hiltner\,600. Some
differences (less than 1\%) were found in the calibrated fluxes of 
the standard star in the overlapping range of the VPHs, which translate to the
flux calibration of the nebular spectra. Nevertheless, these differences are
within the uncertainties of the flux calibration process. Finally, we note
that strong sky lines in the nebular spectrum could not be completely removed,
leaving some residuals that prevent us from a clear identification or
measurement of some nebular emission lines. This problem is particularly
accused in the spectrum obtained with the VPH R2500I where most sky lines could
not be removed (see Sect. 3.2).

\section{RESULTS}
\label{Section:results}

\subsection{Morphology}
The high quality and subarcsecond resolution of the OSIRIS/GTC H$\alpha$ image (Fig.\,2)
allow us to describe new morphological structures in RWT\,152 not detected so
far, improving substantially the previous description by AM15. The nebula shows 
a clear bipolar morphology in the light of H$\alpha$ with a size of $\simeq$ 17$\times$21 arcsec$^{2}$ and 
the major axis oriented at PA $\sim$ 45$^{\circ}$, that are compatible with the size and orientation 
measured by AM15 in their [O\,{\sc iii}] image. However, the new image reveals
that the bipolar lobes are slightly different from each other, being the SW lobe
broader than the NE one and that they are composed
by many bubbles, specially well defined in the SW lobe. No point-symmetric
distribution of the bubbles is noticed in the H$\alpha$ image. The presence of these
small bubbles may explain the deviations from a pure hour-glass geometry
observed in the high-resolution, long-slit spectrum along the major nebular
axis (AM15).

The H$\alpha$ image also shows that the equatorial region of the bipolar shell is
particularly bright (Fig.\,2, more evident in the left panel). To check the veracity of this structure and
to discard possible effects produced by the brightness of the CS, we inspected
the RTF image in the individual band with the optical centre at
$\lambda_{6}$=6600 \AA{} (see Fig.\,3), where the emission should be dominated by the
nebular continuum, although with a small contribution of H$\alpha$ and even much smaller of [N\,{\sc ii}]. 
In this image, the bright equatorial region cannot be recognized, indicating that this region 
corresponds to a real nebular structure that is most probably associated to the
equatorial torus identified in the high-resolution, long-slit spectra (AM15). 
In  Fig.\,2, the CS appears displaced $\sim$ 3\,arcsec northwards with 
respect to the centre of the nebula. This value slightly differs from the
displacement inferred from the high-resolution spectra, where a shift of
$\simeq$ 1.4\,arcsec towards PA $\simeq$ 348$^{\circ}$ was obtained
(AM15). However, as the new image does not allow us to trace clearly the
torus, the displacement obtained from the high-resolution spectra should be
considered as a more precise value.

\begin{figure}
\includegraphics[width=0.48\textwidth]{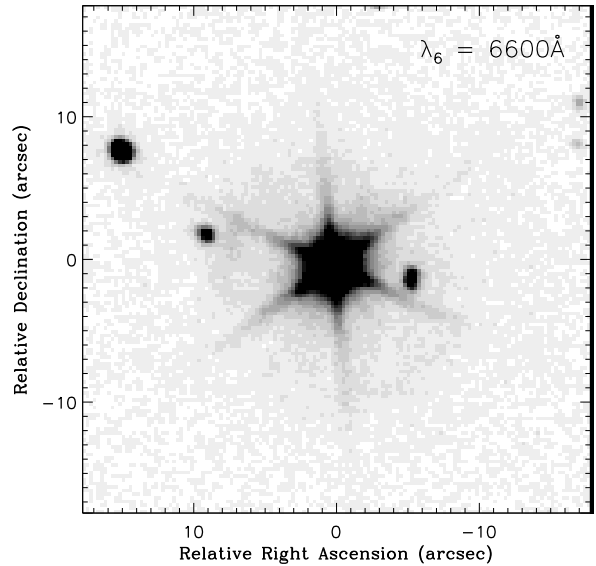}
  \vspace*{1pt}
  \caption{Grey-scale reproduction of the RTF image with the optical centre at $\lambda_{6}$=6600 \AA{}, 
  where the contribution of the H$\alpha$ emission line is minimum (see figure\,1). Grey levels are linear.}
\end{figure}

Finally, a very faint circular halo of $\sim$ 29\,arcsec in diameter can be
recognized in the H$\alpha$ image (see Fig.\,2, right panel). Neither the main
nebula nor the CS are centred in the halo but are clearly displaced towards the
northeast with respect to the geometrical center of the
halo. Displacements of a halo with respect to the CS and PN have been
traditionally attributed to interaction of the PN with the ISM \citep[see, e.g.,][]{Ramos-Larios2009}. 
In the case of RWT\,152, the displacements suggest that the
east/southeast region of the halo is interacting with the ISM. Nevertheless, if
so, one would expect that the east/southeast part of the halo would be the
brightest one, which is not observed in our image. In addition, the halo does
not show departures from a circular symmetry, as it could be expected from that
interaction. Higher-resolution images would be helpful to confirm this
possible interaction.

\subsection{The optical spectrum: physical conditions and chemical abundances}

Figure\,4 shows the intermediate-resolution, long-slit spectra of RWT\,152, both inner nebula (upper panel) 
and halo (lower panel), obtained by combining the VPHs R2500U, R2500V and R2500R. The VPH R2500I is
not shown here because of the strong contamination by the sky lines (Sect. 2.2,
but see also below). For the case of the inner nebula, the spectra have been obtained by integrating the detected emission 
lines between 4 and 6.4 arcsec west of RWT\,152 along the slit. This region 
corresponds to that showing the highest signal-to-noise ratio for the weakest lines (e.g., [O\,{\sc iii}]$\lambda$4363). 
Whereas only H$\alpha$, H$\beta$ and [O\,{\sc iii}]$\lambda$$\lambda$4959,5007 emission lines had been previously detected
in RWT\,152 (AM15), the long-exposure OSIRIS/GTC spectra reveal other
faint nebular emissions. In particular, [Ar\,{\sc iii}], [Ne\,{\sc iii}] and
He\,{\sc i} emission lines are detected. The [N\,{\sc ii}]$\lambda$6584
emission line could also be present, although its extreme faintness (observed
flux $\sim$ 1.32 $\times$10$^{-17}$ erg\,cm$^{-2}$\,s$^{-1}$)
suggests to take this identification with caution. In the VPH R2500I the [S\,{\sc
  iii}]$\lambda$$\lambda$9069,9532 emission lines are clearly identified and
relatively isolated from strong sky lines whereas other faint emission 
lines ([Ar\,{\sc iii}]$\lambda$7751, some Paschen lines) could also be
present. Emission lines due to [N\,{\sc i}], [O\,{\sc i}], [O\,{\sc ii}] and [S\,{\sc ii}] are not
identified. This result is entirely compatible with the lack of these emission lines in the CAFOS 
spectra and allow us to conclude that emission lines from neutral and single-ionized metals are not
present in the spectrum of RWT\,152 (except for the possible [N\,{\sc ii}]$\lambda$6584 emission line). 
Finally, He\,{\sc ii} emission lines are neither identified. 

The spectra of the halo (Fig.\,4, lower panel) have been integrated in two 6.35 arcsec 
regions eastwards and westwards of RWT\,152 along the slit. These regions have been added to obtain a higher S/N. 
Besides the Balmer and [O\,{\sc iii}]$\lambda$$\lambda$4959,5007 emission lines, the spectrum of the halo also shows 
the He\,{\sc i}$\lambda$5876 emission line.

\begin{figure*}
\includegraphics[width=1.\textwidth]{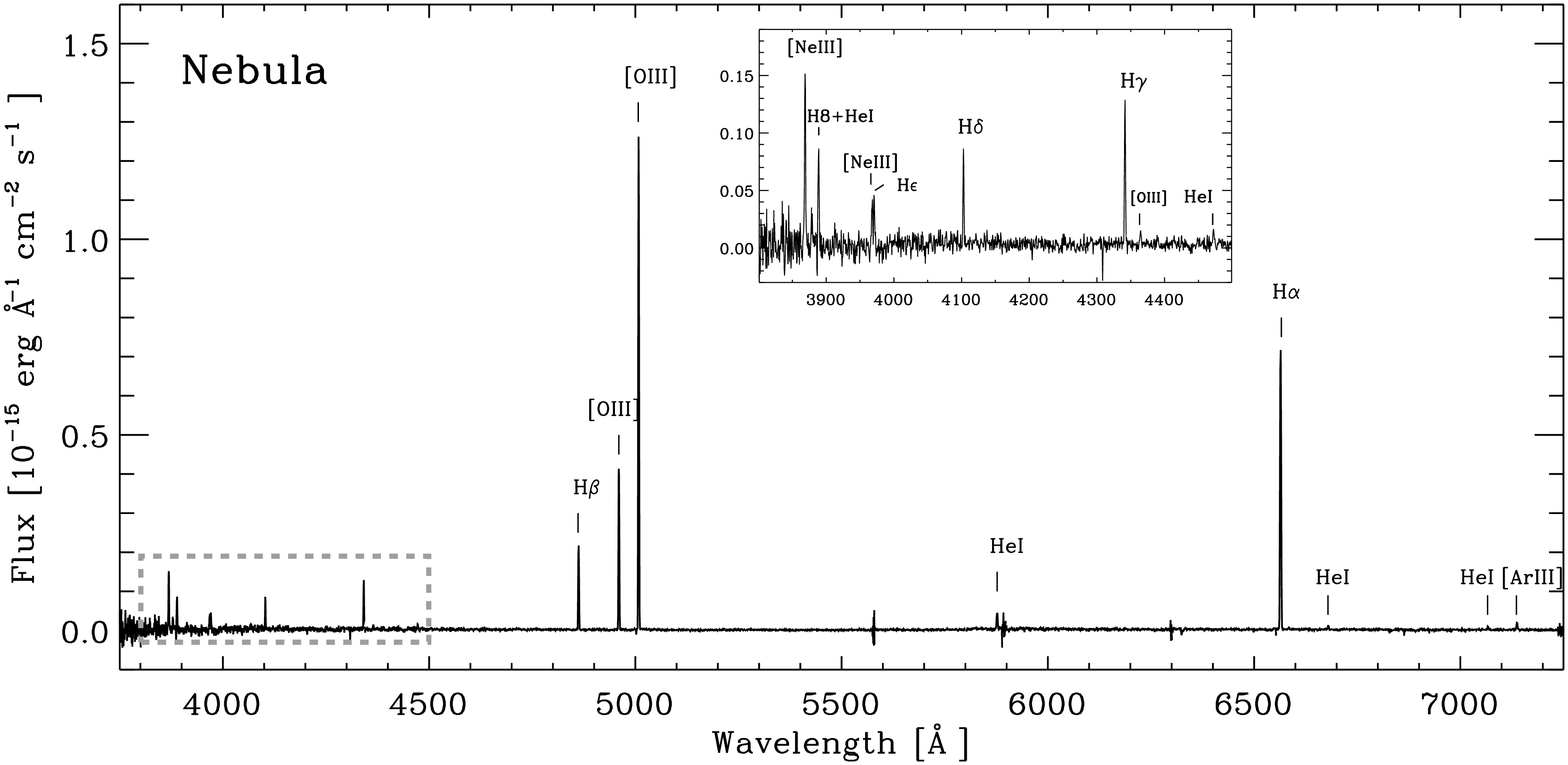}
\includegraphics[width=1.\textwidth]{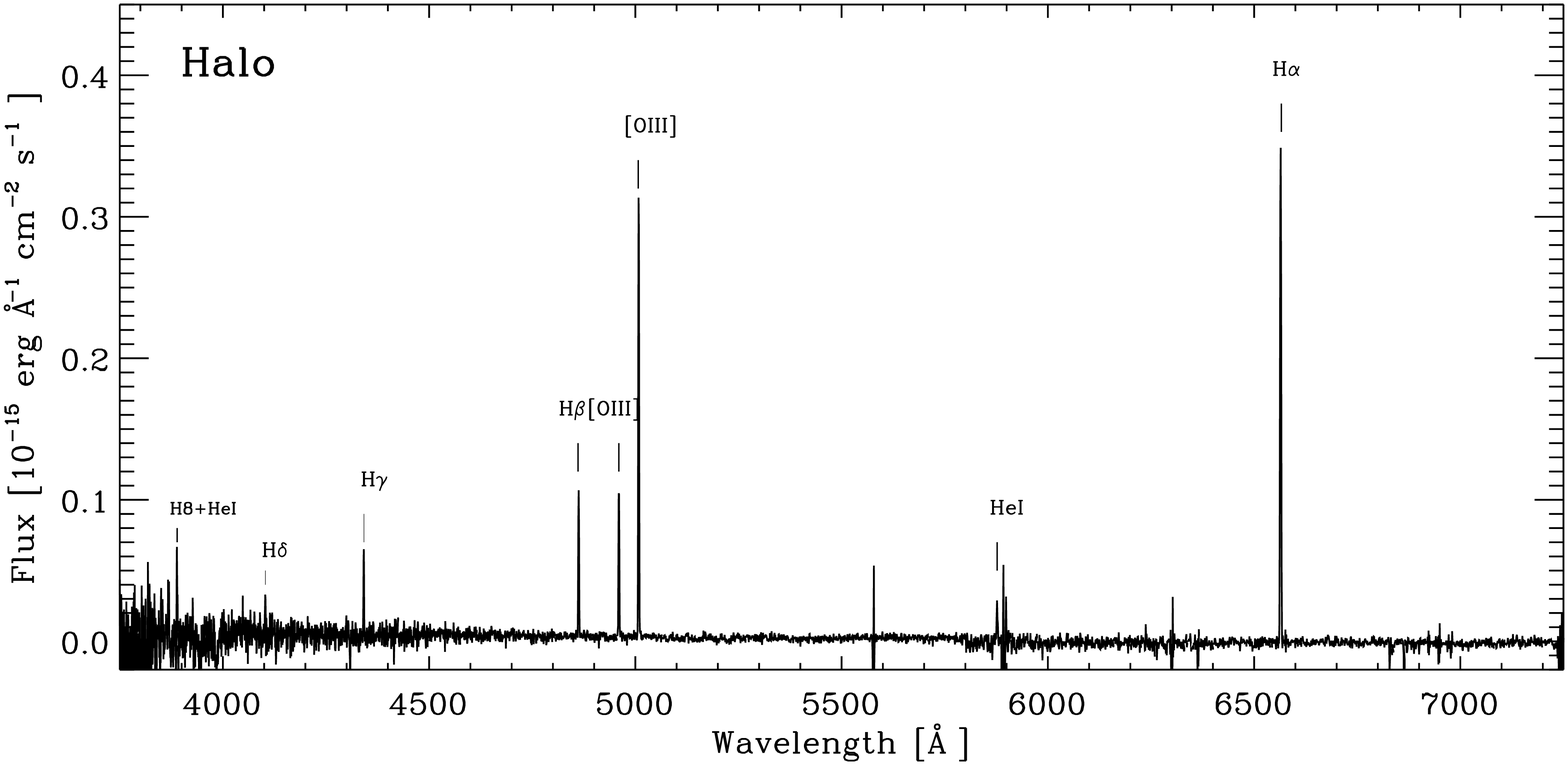}
  \vspace*{1pt}
  \caption{Optical spectrum of the main nebula of RWT\,152 (upper panel) and the halo (lower panel) in 
the spectral range 3800--7250 \AA. The VPHs R2500U, R2500V and R2500R have been combined. The detected
    emissions lines are labelled. The inset in the upper panel shows the spectrum of the inner nebula in the range
    3800--4500 \AA. For the case of the inner nebula, the spectra have been obtained by integrating the detected emission 
lines between 4 and 6.4 arcsec west of RWT\,152 along the slit (see text). The spectra of the halo (lower panel) 
have been integrated in two 6.35 arcsec regions eastwards and westwards of RWT\,152 along the slit.}
\end{figure*}

 \begin{table*}
\caption{Dereddened emission line intensities in RWT\,152.}            
\label{table:1}      
\centering           
\begin{tabular}{lrcccc}   
\hline  
 Line & $f(\lambda)$ &  \multicolumn{2}{c}{Nebula (2.4 arcsec$^{1}$)}	&\multicolumn{2}{c}{Halo (12.70 arcsec$^{1}$)}  \\   
 \cmidrule(l{10pt}r{10pt}){3-4}
 \cmidrule(l{10pt}r{10pt}){5-6}
 &  	& F$_{\lambda}$(observed) 	& I$_{\lambda}$(corrected) & F$_{\lambda}$(observed)  &   I$_{\lambda}$(corrected) \\   
\hline

[Ne\,{\sc iii}] $\lambda$3869          & 0.228    &   52.0 $\pm$ 2.9 &		70.4 $\pm$ 4.3		& 	-- \  & 	-- \\

He\,{\sc i} + H8 $\lambda$3889	& 0.223     &	31.7 $\pm$ 1.8		&  42.7 $\pm$ 2.6   &  46.9 $\pm$ 6.4 &	63.8 $\pm$ 8.7 \\

[Ne\,{\sc iii}] $\lambda$3968          & 0.203    &	9.0 $\pm$ 0.9	    &	11.9 $\pm$ 1.2	    & 	-- \  & 	-- \\

H$\epsilon$ $\lambda$3970          & 0.203     &   14.3 $\pm$ 1.0 &		18.9 $\pm$ 1.4		   & 	-- \  & 	-- \\

H$\delta$ $\lambda$4101          & 0.172    &    24.9 $\pm$ 1.1    &		31.4 $\pm$ 1.4		    & 	30.5 $\pm$ 3.7  &    38.7 $\pm$ 4.8 \\

H$\gamma$ $\lambda$4340          & 0.129    &   40.5 $\pm$ 1.5 &		48.3 $\pm$ 1.9		   & 44.1 $\pm$ 2.2 &  52.8 $\pm$ 2.7 \\

[O\,{\sc iii}] $\lambda$4363    		& 0.124    &   7.6 $\pm$ 0.5&		9.0 $\pm$ 0.6		 &  	-- \  & 	-- \\

He\,{\sc i} $\lambda$4471          &  0.115    &    5.8 $\pm$ 0.3 &		 6.6 $\pm$ 0.4		   & 	-- \ & 	-- \\

H$\beta$ $\lambda$4861          & 0.000    &  100.0 $\pm$ 2.9  &			100.0 $\pm$ 2.9 	    & 100.0 $\pm$ 2.8 & 100.0 $\pm$ 2.8 \\

[O\,{\sc iii}] $\lambda$4959    & $-0.023$    &   194.4 $\pm$ 4.9 &		188.2 $\pm$ 4.7 		 &  104.4 $\pm$ 2.7 & 101.0 $\pm$ 2.6 \\

[O\,{\sc iii}] $\lambda$5007    & $-0.034$     &    579.6 $\pm$ 15.0 &		 552.7 $\pm$ 14.4		&  300.6 $\pm$ 6.8&   286.2 $\pm$ 6.5 \\

He\,{\sc i} $\lambda$5876          & $-0.216$    &   27.0 $\pm$ 0.8 &		20.2 $\pm$ 0.7		    & 28.3 $\pm$ 1.5 & 20.9 $\pm$ 1.2 \\

H$\alpha$ $\lambda$6563         & $-0.323$    &   438.0 $\pm$ 11.3 &		282.5 $\pm$ 11.0		 &   451.6 $\pm$ 10.5 &  287.0 $\pm$ 10.5 \\

[N\,{\sc ii}] $\lambda$6584(?)    & $-0.326$     &	   2.4 $\pm$ 0.1 &		1.2 $\pm$ 0.3	 & 	-- \ & 	-- \\

He\,{\sc i} $\lambda$6678          & $-0.338$     &  5.1 $\pm$ 0.3  &	3.2 $\pm$ 0.3		   & 	-- \  & 	-- \\

He\,{\sc i} $\lambda$7065          & $-0.383$     &  4.8 $\pm$ 0.2  &		 2.8 $\pm$ 0.3		   & 	-- \ & 	-- \\

[Ar\,{\sc iii}] $\lambda$7135    & $-0.391$     &	 11.0 $\pm$ 0.3   &		 6.5 $\pm$ 0.3	 & 	-- \ & 	-- \\

[S\,{\sc iii}] $\lambda$9069    & $-0.606$      &	 6.9 $\pm$ 0.2   &		3.0 $\pm$ 0.2	& 	-- \  & 	-- \\

[S\,{\sc iii}] $\lambda$9532    & $-0.620$      &	 21.8 $\pm$ 0.6   &		 9.2 $\pm$ 0.6	& 	-- \ & 	-- \\

\hline

log$F$$_{\rm H\beta}$(erg\,cm$^{-2}$\,s$^{-1}$) &         &	&	$-15.84$		&   &-16.20\\

\hline    

\multicolumn{4}{l}{$^{1}$ Size of the region where the long-slit spectra were integrated (see text).} \\
                              
\end{tabular}
\end{table*}

The spectra of the inner nebula have been analyzed using the nebular analysis software {\sc Anneb}
\citep{Olguin2011}, which also integrates the {\sc nebular} package of
{\sc iraf/stsdas} \citep{Shaw-Dufour1995}, for deriving physical conditions and
both ionic and elemental abundances. A set of extinction laws are
also included and a proper error propagation is performed. 
 Briefly, {\sc Anneb} obtains the logarithmic extinction coefficient $c$(H$\beta$) and 
the electron temperature from the intensity ratio I(4959+5007)/I(4363) (i.e., $T_{\rm e}$ ([O\,{\sc iii}])), iteratively,
starting with the values derived for $T_{\rm e}$ = 10$^{4}$\,K and the theoretical H$\alpha$/H$\beta$ 
line intensity ratio for case B recombination in the low-density limit \citep{Osterbrock-Ferland2006}; then, 
it derives the dereddened line intensities that are used to calculate the values of $c$(H$\beta$) and $T_{\rm e}$ ([O\,{\sc
  iii}]) again. The process is repeated until $c$(H$\beta$) and $T_{\rm e}$ ([O\,{\sc
  iii}]) converged to the final values. 
  
   A value for the electron density
($N_{\rm e}$) is also necessary for the calculations. Unfortunately, the lack of the [S\,{\sc
  ii}], [Ar\,{\sc iv}] or [Cl\,{\sc iii}] emission lines from RWT\,152 prevents us from deriving  
 $N_{\rm e}$ from the forbidden lines. Therefore, we have used the observed flux in H$\alpha$ (F(H$\alpha$)) to calibrate
   the image and the formulation by \cite{Hua-Kwok1999} to calculate the mean electron density in RWT\,152. 
We considered a total flux corrected for extinction\footnote{For the calculations of $N_{\rm e}$ we have 
used $c$(H$\beta$) = 0.58, calculated from the observed H$\beta$ and H$\alpha$ fluxes, that is virtually identical to 
the value of 0.59 obtained after the iteration procedure with {\sc Anneb}, see below.} of F$_{0}$(H$\alpha$) = 
5.03 $\times$ 10$^{-13}$ erg\,cm$^{-2}$\,s$^{-1}$, observed in a region of 0.8 $\times$ 29 arcsec$^{2}$ (defined by the 
slit width and the angular size of 
the nebula, including the halo), and obtained $N_{\rm e}$ = 91$\times$D[kpc]$^{-1/2}$. For distances of 
2.4\,kpc \citep{Ebbets-Savage1982} and 6.5\,kpc \citep{Pritchet1984}, the electron density is 59 and 
36\,cm$^{-3}$, respectively, with an estimated error of about 10\% in both cases. No particular differences 
were noticed in the resulting parameters and ionic and elemental abundances by using one or other of the 
derived electron density values. We have adopted $N_{\rm e}$ = 50 $\pm$ 10 cm$^{-3}$.

After the iteration procedure in {\sc Anneb}, we obtain $c$(H$\beta$) = 0.59 $\pm$ 0.04 
(calculated from the H$\beta$ and H$\alpha$ fluxes) and $T_{\rm
  e}$([O\,{\sc iii}]) = 14000 $\pm$ 550\,K, that are listed in Table\,1. {\sc Anneb} also provides 
the $c$(H$\beta$) value calculated from other hydrogen lines, as H$\gamma$, for which 
a consistent value of 0.49$\pm$ 0.10 is obtained. Also, the $c$(H$\beta$) obtained for the halo (0.61$\pm$ 0.04) 
is compatible with these results. We note that all these values of $c$(H$\beta$) derived from the OSIRIS spectra 
are slightly higher than (but still compatible with) that obtained by AM15 ($c$(H$\beta$) $\sim$ 0.46), suggesting internal variations of 
the extinction in the nebula. On
the other hand, according to \cite{Schlegel1998}, E(B-V) towards
RWT\,152 is $\sim$ 0.1, implying $c$(H$\beta$) $\sim$ 0.15 (if Rv = 3.1 is assumed), that is very
different from the value of $\sim$ 0.6 derived by us. This large difference would
imply a large amount of internal reddening in the nebula, which would not
be expected for such a low-density nebula evolved from a low-mass progenitor
(see Section 4.1). Alternatively, the difference could be due to the
existence of small scale structure in the Galactic dust distribution towards
RWT\,152, at scales well below $\sim$ 6 arcmin that is the spatial resolution of the
maps by \cite{Schlegel1998}. If such a small scale structure exists, it
would be undetectable in the maps by Schlegel et al. but could be revealed by
observations at much higher spatial resolution than 6 arcmin, as is the case
of our spectra where we can  measured the extinction at spatial scales of
a few arcseconds. In any case, the extinction towards RWT\,152 deserves further
investigation.

Table\,1 lists the observed emission line fluxes and dereddened emission line intensities and
their Poissonian errors, both for nebula and halo spectra, where the emission line fluxes have been dereddened with 
$c$(H$\beta$) = 0.59 and the extinction 
law $f$($\lambda$) of \cite{Seaton1979}. We further note that the use of other extinction laws \citep[e.g.,][]{Cardelli1989} does not produce appreciable differences in the dereddened emission 
line intensities and other parameters.

 \begin{table}
\caption{Mean ionic abundances relative to H$^{+}$ of RWT\,152.}            
\label{table:1}      
\centering           
\begin{tabular}{lcc}   
\hline
                     
Ion & Ionic abundance\\   
\hline    
He$^{+}$	&	0.139$\pm$0.004\\
O$^{++}$	&	(7.0$\pm$0.1)$\times$10$^{-5}$\\
N$^{+}$	&	(1.3$\pm$0.1)$\times$10$^{-7}$\\
Ar$^{++}$	&	(2.9$\pm$0.1)$\times$10$^{-7}$\\
Ne$^{++}$	&	(1.8$\pm$0.1)$\times$10$^{-5}$ \\
S$^{++}$	&	(2.9$\pm$0.1)$\times$10$^{-7}$ \\
\hline
\end{tabular}
\end{table}

The absence of the He\,{\sc ii}$\lambda$4686 emission line and the [O\,{\sc
  iii}]/H$\beta$ line intensity ratio of $\simeq$ 8 (Table\,1) indicate a very
low-excitation PN, with an excitation class of 2 according to \cite{Gurzadian-Egikian1991}. 
We note that RWT\,152 was erroneously classified as a relatively high-excitation
PN by AM15. The very low-excitation is compatible with the relatively low
effective temperature of the CS of RWT\,152 ($\simeq$ 45000\,K, \citealt{Ebbets-Savage1982}). 
In these circumstances, the non-detection of neutral and single-ionized
emission lines (e.g., [S\,{\sc ii}], [O\,{\sc i}], [N\,{\sc i}], [O\,{\sc
  ii}]) and the extremely faintness of the (possible) [N\,{\sc
  ii}]$\lambda$6584 emission line is highly peculiar for this excitation class.
  Under these conditions, one could expect that emission lines due to,
e.g., O$^{0}$, O$^{+}$, S$^{+}$, N$^{0}$, were prominent even though the
abundances of O, S, N were low. This spectrum can be understood if RWT\,152 is
a density-bounded PN with no low-excitation region (at least along the bipolar
lobes, because the equatorial structure has not been studied). In fact, [O\,{\sc i}], [N\,{\sc ii}],
[S\,{\sc ii}] emission lines become very weak in density-bounded models
\citep[see][]{Gesicki-Zijlstra2003}, and, in addition, a PN becomes optically thin
if the effective temperature of the CS is in the range 40000--50000\,K \citep{Kaler-Jacoby1991}, 
as it is the case of RWT\,152.

 \begin{table}
\caption[]{Elemental abundances of RWT\,152$^{1}$ and average abundances
  for type III and type IV PNe taken from \cite{Costa1996}.}            
\label{table:1}      
\centering           
\begin{tabular}{lccc}   
\hline           
Element & Abundance & Type III PNe             & Type IV PNe \\   
\hline    
He/H		        & 0.139$\pm$0.004      & 0.099 & 0.104	\\
$\epsilon$(O/H)	        & 7.85$\pm$0.02	       & 8.42  & 8.08	\\
$\epsilon$(N/H) 	&  5.11$\pm$0.03       & 7.74  & 7.41	\\
$\epsilon$(Ar/H)	&  5.73$\pm$0.02       & 6.07  & 5.22	\\
$\epsilon$(Ne/H) 	&  7.26$\pm$0.03       & 7.71  & 7.27	\\
$\epsilon$(S/H) 	&  5.52$\pm$0.08       & 6.74  & 5.64   \\
\hline
\multicolumn{4}{l}{$^{1}$ See text for how the elemental abundances have been obtained.} \\

\end{tabular}
\end{table}

The derived ionic abundances (i.e., the number density ratio relative to the H$^{+}$ number density), calculated as a weighted average by the
signal-to-noise ratio of each line for species with more than one line, are listed
in Table\,2. To obtain the helium abundance, we used the method by \cite{Kwitter-Henry2001}, 
while for the argon abundance, we followed the method by \cite{Kingsburgh-Barlow1994} and used an icf = 1.87, since no [Ar\,{\sc ii}] is observed. These helium and argon elemental abundances are listed in Table\,3, where
$\epsilon({\rm X/H}) = \log({\rm X/H})+12$ is given (being 12 the hydrogen number density).

 For the rest of the elements, the absence of some emission lines prevents to calculate the icfs and, 
therefore, the elemental abundances. Nevertheless, the low effective temperature of the CS and
the observed spectrum allow us to make some reasonable assumptions to obtain
approximate values for the abundances. In particular, the lack of
He\,{\sc ii} and [Ar\,{\sc iv}] emission lines in the spectra makes it highly
improbable that ionization states as O$^{3+}$ and Ne$^{3+}$ may exist
in the nebula. This, and the absence of emission lines from  O$^0$, O$^{+}$, and Ne$^{+}$ in the spectrum 
of RWT\,152 strongly suggest that O$^{2+}$ and Ne$^{2+}$ are the dominant excitation 
states. In consequence, their ionic abundances may be considered as representative of their elemental
abundances and those are listed in Table\,3.

In the case of sulfur, if we assume that the ionic abundance of S$^{2+}$
is representative of the elemental sulfur abundance, the elemental
abundance would be 5.46. However, this assumption is probably erroneous
for sulfur because even if He\,{\sc ii} and [Ar\,{\sc iv}] emission lines
are unseen, faint [S\,{\sc iv}] emission lines, as, e.g., at 10.52 $\mu$m,
may be detected in infrared spectra of some low-excitation PNe with
relatively cool CSs. In fact, the existence of O$^{2+}$ (ionization
potential = 35.1\,eV) suggests that S$^{3+}$ (ionization potential =
34.8\,eV) may also exists in RWT\,152. This is the case of K\,648, a very
similar object to RWT\,152, with an sdO central star of $T_{\rm eff}$ =
37000\,K \cite{Otsuka2015,Heber1993,Bianchi2001}. As far as we know, there
is no available infrared spectrum of RWT\,152. Therefore, for comparison
purposes, we have estimated the contribution of [S\,{\sc iv}] to the total
abundance of sulfur in the case of K\,648 with the data provided by
\cite{Otsuka2015}. Our calculations show that the elemental abundance of
sulfur reported by \cite{Otsuka2015} is reduced by $\sim$ 15\% if the
contribution of the [S\,{\sc iv}]$\lambda$10.52 $\mu$m emission line is
not considered. If we assume a similar contribution of a (possible)
[S\,{\sc iv}]$\lambda$10.52 $\mu$m emission line to the total abundance of
sulfur in RWT\,152, the value of 5.46 would increase to 5.52. From these
calculations and taking into account the differences between K\,648 and
RWT\,152, we will consider in Table\,3 a value for the elemental sulfur
abundance in RWT\,152 of 5.52 $\pm$ 0.08, where the error is probably an
upper limit to the actual error.

\begin{figure}
\includegraphics[width=0.48\textwidth]{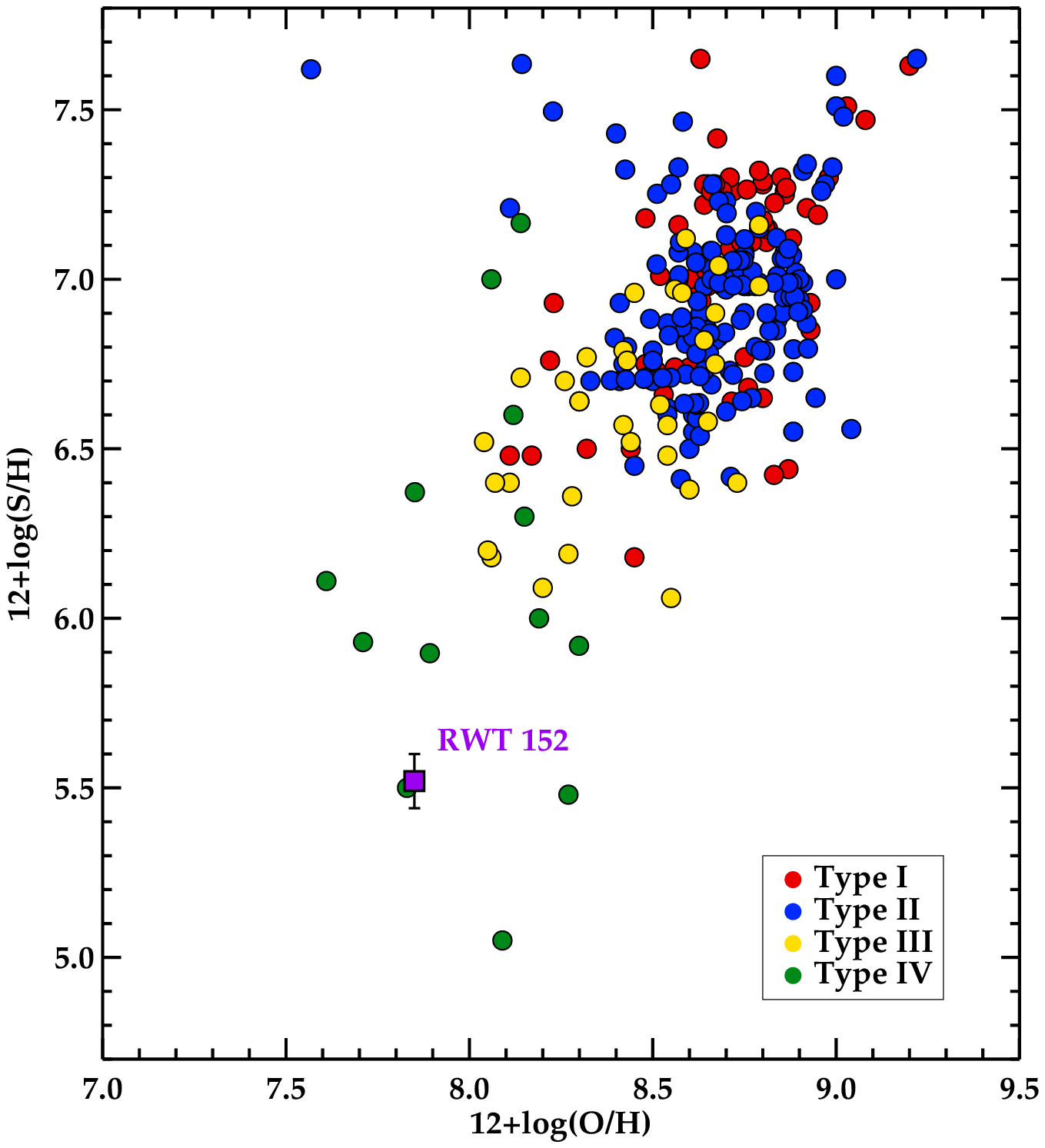}
  \vspace*{1pt}
  \caption[]{Plot of 12+log(Ar/H) versus 12+log(O/H) (left panel) and 12+log(S/H) versus 12+log(O/H) (right panel) 
    for a sample of type I (red), type II (blue), type III (yellow) and type IV (green) PNe. Data from \cite{Henry2004}, \cite{Costa1996}, \cite{Maciel-Koppen1994}, \cite{Howard1997}, and \cite{Pereira-Miranda2007}. The position of RWT\,152 is marked with a purple square.}
\end{figure}

Finally, the nitrogen abundance in RWT\,152 is the most uncertain one. 
We note that N$^{2+}$ may exist in the nebula since other especies with higher ionization potential, like O$^{2+}$, are also detected. We have inspected the International Ultraviolet Explorer (IUE) spectra of RWT\,152 to search for the presence of the semi-forbidden N\,{\sc iii}]\,$\lambda$$\lambda$1747,1754 emission lines. However, there is no traces of these emission lines 
 although they should be prominent if the abundance of N is within the range of values observed in PNe.
 Therefore, it seems that N$^{2+}$ does not exist in the nebula. Taking
this into account and that emission lines due to N$^0$ neither exist, the N abundance 
should be similar to the abundance of  N$^{+}$ and this is quote in Table\,3.
Nevertheless, given the extreme faintness of the
[N\,{\sc ii}]$\lambda$6584 (see above), the existence of nitrogen in the
nebula is questionable and, in any case, the elemental abundance in Table\,3 should
be considered as an upper limit. This result is very surprising for a PN.
 Nitrogen may not have been produced during the
evolution of RWT\,152 by the second dredged up. However, this mechanism
alone hardly explains the extremely low N abundance or, even, its absence
in RWT\,152. As additional possibilities we may consider that RWT\,152 has been formed in
a very nitrogen-poor environment and/or that nitrogen is depleted (forming part
of grains). In any case, a reasonable explanation has not been found.

Table\,3 shows low elemental abundances of metals, confirming the previous
suggestion by AM15. We will discuss below the implications of these
elemental abundances. It is worth noting that the low metal abundances
found in RWT\,152 provides strong support for a deficiency in metals in
2M\,1931+4324, another PN+sdO system with a nebular spectrum very similar
to that of RWT\,152 \citep[][see also above]{Aller2013}.

\section{DISCUSSION}
\label{Section:discussion}

\subsection{Chemistry and evolutionary status of RWT\,152}
 To obtain information about the progenitor of the CS of RWT\,152, it
is interesting to classify RWT\,152 according to the \cite{Peimbert1978} types. \cite{Peimbert1978}
 classified PNe into four types on the basis of 
their chemical composition, spatial distribution and kinematical properties. Type\,I 
PNe are those with high He and N (He/H $\ge$ 0.125; log(N/O) $\ge$ -0.30, see \cite{Peimbert-Torres1983}) and
they are the youngest
population. Type\,II PNe are the intermediate population, generally older 
than type\,I PNe and, therefore, more deficient in heavy elements and without
He and N enrichments. Type\,III PNe belong to the thick disk and present
peculiar velocities of $\mid \Delta V_{pr}\mid \ge$ 60 km\,s$^{-1}$, 
which is the difference between the observed radial velocity and that expected
on the basis of Galactic rotation (assumed circular). Finally, type\,IV PNe 
are halo objects with $\mid \Delta V_{pr}\mid \ge$ 60 km\,s$^{-1}$ and log
(O/H)+12 $\le$ 8.1; they correspond to the oldest population. 

The helium abundance derived for RWT\,152 (He/H $\sim$ 0.140, Table\,3) indicates a
 type\,I or type\,II PN. However, the extremely low (or absent) nitrogen
abundance rules out these two types. A comparison of the abundances of O, S, Ar and Ne 
with those typical of type\,III and type\,IV PNe, shown in Table\,3, indicates that RWT\,152 is a type\,IV PN.
 To reinforce this classification, we show in Figure\,5 the 12+log(S/H) versus 12+log(O/H) 
diagram for the four Peimbert types of PNe, in which RWT\,152 is placed in the region of halo PNe.
Finally, it is worth mentioning that the helium abundance in RWT\,152 appears high for type\,IV PNe, 
although some halo PNe do present relatively high helium
abundances, as, e.g., M\,2-7 with He/H $\sim$ 0.137 \citep{Quireza2007}.

Moreover, we have derived 
the peculiar velocity of RWT\,152 from its heliocentric radial velocity 
$V_{\rm HEL}$ $\sim$ +134.5$\pm$1.8\,km\,s$^{-1}$ (AM15), following the
formulation explained in \cite{Pena2013}, and the resulting value is 
$\sim$ 92 - 131 km\,s$^{-1}$ for distances of 2.4\,kpc
and 6.5\,kpc, respectively, in any case $\ge$ 60 km\,s$^{-1}$ (see above), confirming the type\,IV classification for RWT\,152. Finally, 
we note that the height above the Galactic plane of RWT\,152 is 0.2--0.8\,kpc for the mentioned
distances. Although type\,IV PNe are usually located at z $>$ 0.8 \citep{Peimbert1990}, other halo PNe are 
located at comparable heights above that plane \citep[see][and references therein]{Pereira-Miranda2007}. 
Summarizing, the chemical abundances and the peculiar velocity add RWT\,152 as a new member of the few known halo PNe.

A comparison of the He and O abundances in RWT\,152 with evolutionary models
of stellar yields by \cite{Marigo2001} suggests a progenitor star with an initial mass
of $\sim$ 1.3\,M$_{\odot}$ and very low metallicity Z = 0.004. The stellar
mass is compatible with that expected for sdOs \citep{Heber2009} and the low
metallicity indicates that the progenitor was formed in a poor-metal
environment. In contrast, similar models of initial mass
$\sim$ 1.25\,M$_{\odot}$ and Z = 0.004 by \cite{Karakas2010} predict abundances substantially different 
from those found in RWT\,152. Even for the lowest metallicity in the Karakas models (Z = 0.0001), we do not recover 
the chemical abundances of RWT\,152. Therefore, it is clear that drawing conclusions about the progenitor star of RWT\,152 from the chemical abundances 
of the nebula may be a bit misleading. For this reason, we have obtained information about the progenitor from the current 
status of its CS. Figure\,6 shows the position of RWT\,152 
in the HR diagram $\log g -  T_{\rm eff}$ with the post-AGB tracks by \cite{Bloecker1995} and \cite{Schonberner1983}.
  The location of the star \citep[$T_{\rm eff}$ $\simeq$ 45\,000\,K, $\log g$ $\simeq$ 4.5;][]{Ebbets-Savage1982} is consistent with a current mass of
    M$\sim$0.55 ${\rm M}_{\sun}$ which implies an initial mass in the main sequence of $\sim$ 1 ${\rm M}_{\sun}$. 
 For such a low-mass star, the ejected mass during the AGB evolution is expected to be 
small. In fact, with the electron density derived above and the size of the nebula we obtain values of 
1.3$\times$10$^{-2}$ - 1.6$\times$10$^{-1}$ M$_{\odot}$ for the ionized nebular mass (assuming 2.4 and 6.5\,kpc, 
respectively, and a filling factor of 0.6). These values are much smaller than ionized masses usually obtained for 
PNe \citep[see, e.g.,][]{Hua-Kwok1999}, further supporting a low-mass progenitor for RWT\,152. Moreover, taking into 
account that 0.1-0.3 M$_{\odot}$ are lost in the RGB phase of a low-mass star \citep{Dorman1993}, the current 
mass of the CS and the obtained ionized mass, we recover a  progenitor star with a $\sim$ 0.8-1.0 M$_{\odot}$, 
 in agreement with the value obtained from the position of the CS in the $\log g -  T_{\rm eff}$ diagram. 
 Such a low ionized mass, combined with a relatively high kinematical age, may explain the 
 low surface brightness of RWT\,152. If a low-mass progenitor is involved in the evolution of other 
PNe+sdO systems, it is not surprising that these PNe are very faint and, in some cases, may have faded
beyond detection.

\begin{figure}
\includegraphics[width=0.49\textwidth]{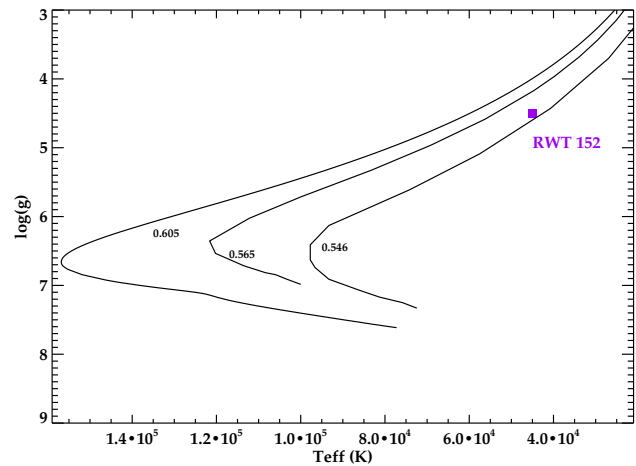}
  \vspace*{1pt}
  \caption[]{Position of RWT\,152 in the $\log g -  T_{\rm eff}$ diagram according to \cite{Ebbets-Savage1982}. Post-AGB by evolutionary tracks by \cite{Bloecker1995} and \cite{Schonberner1983} are drawn and labelled with the corresponding stellar mass (in M$_{\odot}$).}
\end{figure}

\subsection{The morphology of RWT\,152}
The GTC image shows that RWT\,152 presents multiple bubbles in its bipolar
lobes. The number of PNe with multiple bubbles (or lobes) has largely increased,
mainly due to increasing resolution of imaging capabilities. PNe with multiple
bubbles/lobes can be classified in two broad categories: (1) those with a
point-symmetric distribution of the bubbles/lobes as, e.g., NGC\,6058 \citep{Guillen2013}, starfish-like PNe \citep{Sahai2000}, 
and other young PNe \citep{Sahai-Trauger1998}; and (2) those with a random (non-point-symmetric) distribution of the
bubbles/lobes as, e.g., NGC\,1514 \citep{Ressler2010}, NGC\,7094 and
Abell\,43 \citep{Rauch1999}. RWT\,152 seems to be an intermediate case: it shares
with NGC\,1514 and Abell\,43 the random distribution of the bubbles but also shares with starfish
PNe the existence of a bright equatorial region (torus) separating two main lobes (or multiple point-symmetric lobes). 

Several scenarios 
have been proposed to explain the presence of multiple bubbles/lobes, including bipolar
jets at several directions with variable ejection velocity \citep[e.g.,][]{Velazquez2012}, interaction of a fast wind with a warped circumstellar disk
\citep{Rijkhorst2005}, and interaction of a fast wind with a
inhomogenous spherical shell \citep{Steffen2013}. Multiple point-symmetric
bubbles/lobes appear more compatible with a bipolar jet model or warped disk
model than with an inhomogenous shell scenario; the latter would require an extremely point-symmetric
density distribution in the spherical AGB shell, which could be difficult to explain. PNe with
randomly distributed bubbles could be better explain by an inhomogenous shell. In the
case of RWT\,152, the random distribution of the bubbles points out to a model
in which a dense equatorial torus exists in the spherical shell while the rest of the shell presents an inhomogenous density distribution. Alternatively, wind interaction with a warped disk could
also explain random bubbles if material in the two sides of the disk is distributed in a inhomogenous
manner. Nevertheless, a definitive conclusion about the shaping of RWT\,152 is difficult to be reached because it is 
an already evolved PNe and the original shaping mechanism could be masked by other processes (e.g., hydrodynamical 
instabilities). Images of RWT\,152 at higher spatial resolution and, in particular, of its equatorial region 
would be very useful to complete description of the nebula.

\section{Conclusions}
\label{Section:conclusions}

We have presented OSIRIS/GTC red tunable filter H$\alpha$ imaging
and intermediate-resolution, long-slit spectroscopy of RWT\,152, one of the few known 
PNe hosting an sdO central star. The
data, obtained at subarcsec spatial resolution, allowed us to describe the
detailed morphology and to obtain the physical conditions and chemical abundances of
the nebula. The main conclusions of this work can be summarized as follows:

\begin{enumerate}[(1)]

\item The new H$\alpha$ image shows that RWT\,152 is a bipolar PN with a
  bright equatorial torus, surrounded by a circular halo. The bipolar lobes
  consist of multiple bubbles with a non point-symmetric distribution. The
  center of the halo does not coincide with the center of nebula and with the
  central star, suggesting interaction of the halo with the interstellar
  medium.

 \item The nebular spectra reveal very faint [Ar\,{\sc iii}], [Ne\,{\sc iii}],
   He\,{\sc i}, [S\,{\sc iii}] emission lines, which had not been previously detected. Emission lines due
   to neutral and single-ionized metals are not detected, except for an extremely 
faint [N\,{\sc ii}]$\lambda$6584 emission line, while the lack of He\,{\sc ii}
   emission lines indicates that high ionization states (e.g.,
   O$^{3+}$, S$^{3+}$, N$^{3+}$) are probably not present in the nebula. These
   results strongly suggests that RWT\,152 is a density-bounded PN, at least
   in the direction of the bipolar lobes.

 \item An electron temperature $T_{\rm e}$([O\,{\sc iii}]) of $\sim$ 14000\,K
and a very low electron density $N_{\rm e}$ $\sim$ 50 cm$^{-3}$ have been obtained
for the nebula.

 \item Except for helium, the derived chemical abundances of S, O, Ne and Ar are low. 
 The abundance of N seems to be extremely low and an 
explanation for this fact has not been found yet. The low abundances and the high peculiar velocity
  of the object ($\mid \Delta V_{pr}\mid \sim$ 92-131 km\,s$^{-1}$) strongly suggest that RWT\,152 is a halo PN,
  adding a new member to the small number of known PNe in this type.

 \item A comparison of the atmospheric parameters of the CS with post-AGB evolutionary tracks suggests
a $\sim$ 1.0 M$_{\odot}$ progenitor for RWT\,152, that should be formed in a metal-poor environment. This low-mass progenitor is 
compatible with the small ionized nebular mass (1.3$\times$10$^{-2}$ -- 1.6$\times$10$^{-1}$ M$_{\odot}$) obtained for RWT\,152.

\item The multiple, non-point-symmetric bubbles observed in the bipolar lobes
  could be attributed to interaction of a fast wind with an inhomogeneous
  distribution of material at both sides of the equatorial
  torus. Nevertheless, as RWT\,152 is an already evolved PN, the original
  mechanism for the formation of the multiple bubbles could be masked by other 
phenomena that have occurred through the evolution.

\end{enumerate}

\section*{Acknowledgments}

This paper has been partially supported by grant AYA\,2011-24052 (AA, ES), 
AYA\,2011-30228-C3-01 (LFM), ESP2014-55996-C2-2-R (AU), and AYA2014-57369-C3-3-P (LFM)
of the Spanish MINECO (all them co-funded by FEDER funds). We also acknowledge
support from grant INCITE09\,312191PR (AU, LFM, AA) of Xunta de Galicia, 
partially funded by FEDER funds, from grant PRX15/00511 (AU) of the Spanish MECD, 
and from grant 12VI20 (AU, LFM, AA) of the University of
Vigo. AA also acknowledges support from FONDECYT through postdoctoral grant  3160364.
Authors are very grateful to the staff on the El Roque de los Muchachos
Observatory and specially to Antonio Cabrera for helping us to plan and
reduce the observations successfully. We also thank Irene Pintos for her
detailed explanations about the tunable filters. We acknowledge support from the Faculty of the European
 Space Astronomy Centre (ESAC). This research has made use of
the SIMBAD database, operated at the CDS, Strasbourg (France), Aladin, NASA's
Astrophysics Data System Bibliographic Services, and the Spanish Virtual
Observatory supported from the Spanish MINECO through grant AYA2011-24052.


\bibliographystyle{mn2e} 
\bibliography{Bibliography} 

\label{lastpage}

\end{document}